\documentstyle[aps,prb,epsfig,multicol,amsmath]{revtex}

\begin{document}

\date{13 June 2000}
\draft

\title{Thermoelectric properties of junctions between metal
and strongly correlated semiconductor}

\author{Massimo Rontani}

\address{INFM and Dipartimento di Fisica, Universit\`a degli Studi di Modena
e Reggio Emilia, 41100 Italy,\\ and Department of Physics, 
University of California San Diego,
La Jolla, California 92093-0319}

\author{L.~J.~Sham}

\address{Department of Physics, University of California San Diego,
La Jolla, California 92093-0319}

\maketitle

\begin{abstract}
We propose a junction of metal and
rare-earth compound semiconductor
as the basis
for a possible efficient low-temperature thermoelectric device.
If an overlayer of rare earth atoms differing from the bulk is placed
at the interface, very high values of the figure of merit $ZT$ can be
reached at low temperature. This is due to sharp
variation
of the transmission coefficient of carriers across the junction
at a narrow energy range,
which is intrinsically linked to the localized character
of the overlayer $f$-orbital.
\end{abstract}

%\pacs{72.10.Fk 73.40.Cg 73.40.Ns 73.50.Lw }

\begin{multicols}{2}
\narrowtext

New thermoelectric coolers and power generators are
attracting increasing interest.\cite{physicstoday}
Their quality is
governed by the dimensionless figure of merit $ZT$,
a product of transport coefficients to be defined below
times the absolute temperature. At present,
the best $ZT$ obtainable is $\sim$ 1 at room
temperature, while a value about 3 $\sim$ 4 is
considered a real breakthrough.

There are two theoretical ideas for increasing $ZT$.
One is to utilize the sharp energy features in bulk materials
such as strongly correlated semiconductors.\cite{procnat,mao}
Another is to exploit
the interface energy structures\cite{edwards} in metal/superconductor
junctions, which have been tested experimentally.\cite{nahum}

In this Letter, we combine these two ideas in exploring the thermopower
behavior of a junction between a metal and a particular class of
mixed-valent semiconductors with high dielectric constants, such as
SmB$_6$ and Sm$_2$Se$_3$, called
the electronic ferroelectrics (FE).\cite{FE}
Various cases of the junction are considered,
including clean and ``dirty'' interfaces.
When a suitable rare-earth impurity
layer forms the interface, we find
very high values of $ZT$ at low temperatures.
There is great interest in low temperature thermoelectrics for special
applications.

\begin{figure}
\centerline{\epsfig{file=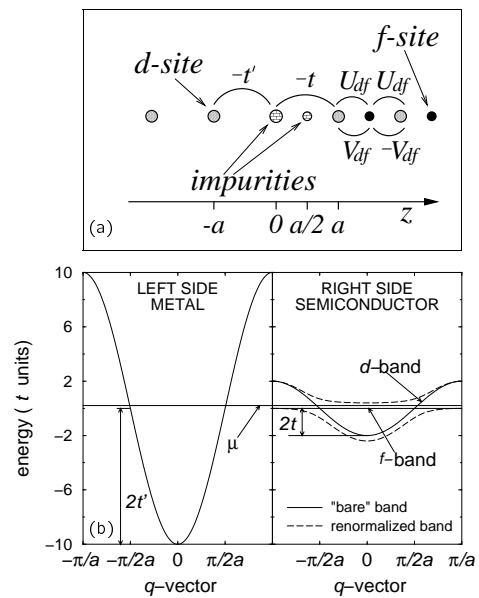,width=2.7in,,angle=0}}
\caption{
\label{fig1}
(a) Pictorial representation of the junction along the $z$ axis perpendicular
to the interface.
(b) ``Bare'' band structure  (\protect{$\Delta=0$})
of materials on both sides of the junction (solid lines), and
renormalized
bands (dashed lines) at $T=0$. Parameters are $t^{\prime}=5t$ and
\protect{$\Delta\!\left(T=0\right)=0.49t$}.
}
\end{figure}

Mixed-valent semiconductors are rare-earth compounds, usually cubic,
whose relevant electronic properties may be modeled by a
$f$-flat band and a broad conduction band, with two electrons
per unit cell.\cite{kondo}
Strong correlation between $d$-electrons
and $f$-holes can renormalize the bands and create a
temperature dependent small gap [see Fig.~\ref{fig1}(b)]:
we consider this case for
the semiconductor on one side of the junction,
and describe it with the self-consistent mean field solution (MF)
of the Falicov-Kimball model.\cite{Falicov}
The ground state of the insulating phase is found to be a coherent condensate
of $d$-electron and $f$-hole pairs, giving a net built-in macroscopic
polarization which breaks the crystal inversion symmetry and makes
the material ferroelectric.\cite{FE}

While previous calculations have dealt with
metal/ordinary-semiconductor junctions neglecting the dependence of
the transmission coefficient on the carrier energy,\cite{thermionic,moyzhes}
for the metal/FE junction this feature turns out to be crucial,
requiring a careful treatment. 
\noindent To this aim, we model the motion along
the $z$ direction perpendicular to the interface with
the one dimensional spinless hamiltonian $\cal{H}$, given by the sum of
three terms describing the left bulk metal, the interface
layer, and the right bulk FE, respectively [see Fig.~\ref{fig1}(a)]:
\begin{equation}
{\cal{H}}
={\cal{H}}_{\rm metal}+{\cal{H}}_{\rm interface}+{\cal{H}}_{\rm FE}.
\label{eq:htot}
\end{equation}
${\cal{H}}_{\rm metal}$ is a tight-binding hamiltonian
for lattice sites at $z=aj$, $j<0$
($a=$ lattice constant), with energy
$\varepsilon_d^{\prime}$ and hopping
parameter $t$.
\protect{${\cal{H}}_{\rm FE}$}
is the bulk Falicov-Kimball hamiltonian:
\begin{eqnarray}
{\cal{H}}_{\rm FE}&=&\tilde{\varepsilon}_d
\sum_{j>0}d^{\dagger}_j d_j
 +\tilde{\varepsilon}_f
\sum_{j>0} f^{\dagger}_{j+1/2} f_{j+1/2}
-t\sum_{j\ge 0} d^{\dagger}_j d_{j+1}
\nonumber\\
-V_{df}&&\!\!\!\!\!\!\!\!\sum_{j\ge 0} f^{\dagger}_{j+1/2}d_j
+ V_{df}\sum_{j>0}f^{\dagger}_{j-1/2}d_j
{\rm \;+\; H.c.}\nonumber\\
+U_{df}&&\!\!\!\!\!\!\!\!
\sum_{j\ge 0}d^{\dagger}_j d_j
f^{\dagger}_{j+1/2} f_{j+1/2} +U_{df}\sum_{j>0}
d^{\dagger}_j d_j f^{\dagger}_{j-1/2} f_{j-1/2}.
\label{eq:hfe}
\end{eqnarray}
Here,
while electrons can tunnel between $d$-sites
with hopping coefficient $t\ll t^{\prime}$,
$f$-electrons
are completely localized [$f_{j+1/2}$ ($d_j$)
destroys a $f$- ($d$-)electron at
$z=(j+1/2)a$ ($ja$) with energy $\tilde{\varepsilon}_f$
($\tilde{\varepsilon}_d$)],
but interact with $d$-electrons
via the Coulomb repulsion energy $U_{df}$.
Moreover,
the $d$-$f$ hybridization $V_{df}$ is odd under inversion.
On this side of the junction we assume
one electron per unit cell.
The hamiltonian term
\begin{equation}
{\cal{H}}_{\rm interface}=\tilde{\varepsilon}_{d0}d^{\dagger}_0 d_0+
\tilde{\varepsilon}_{f1/2} f^{\dagger}_{1/2}f_{1/2}
\label{eq:twoint}
\end{equation}
accounts for the possibility of impurity atomic layers at the interface,
namely one at $z=0$ made of $d$-like
sites with energy $\tilde{\varepsilon}_{d0}$
and another one at $z=a/2$ made of $f$-sites with localized orbitals of energy
$\tilde{\varepsilon}_{f1/2}$.
In our model, the electrostatic
potential $V$ across the junction (by applying an external
bias) is constant
in the two bulk regions (the small gap FE semiconductor
has carriers at finite temperature) and
has a sharp step at the interface.
We adopt the following strategy to solve the motion across the junction:
(i) We apply the MF approximation to the hamiltonian \protect{${\cal{H}}
\rightarrow {\cal{H}}^{\rm MF}$}, introducing
the built-in coherence of the FE bulk
\protect{$\Delta=U_{df}<d^{\dagger}_j f_{j\pm 1/2}>$}
(the vacuum is the state with no $f$-holes),\cite{FE}
treating renormalized site energies
(\protect{$\tilde{\varepsilon}_d\rightarrow \varepsilon_d $},
\protect{$\tilde{\varepsilon}_f\rightarrow \varepsilon_f $})
as material parameters, and choosing
\protect{$\varepsilon^{\prime}_d=\varepsilon_d=\varepsilon_f=0$}, namely the
middle of the $d$-band on both sides of the junction and the flat $f$-band
are aligned [see Fig.~\ref{fig1}(b)].
(ii) We calculate the site coefficients
for the Bogoliubov-Valatin operator\cite{Degennes}
\begin{equation}
\gamma^{\dagger}_{ke}=\frac{1}{\sqrt{N_s}}\sum_j \left\{
u_k\!\left(j\right)d^{\dagger}_j + v_k\!\left(j+1/2\right)
f^{\dagger}_{j+1/2} \right\}
\label{eq:direct}
\end{equation}
($N_s=$ number of $d$-sites)
which creates elementary electronic excitations of
energy $\omega\!\left(k\right)>0$ (referred to the chemical potential $\mu$)
if applied
to the ground state, i.e.
\begin{equation}
{\rm i}\hbar\dot{\gamma}_{ke}=\left[\gamma_{ke},{\cal{H}}^{\rm MF}
\!-\mu N\right]=
\omega\!\left(k\right)\gamma_{ke},
\label{eq:key}
\end{equation}
where $N$ is the number operator,
and $k$ is a suitable quantum number.\cite{clarification}
Similar expressions occur for holes. From Eqs.~(\ref{eq:direct})
and (\ref{eq:key}) one derives
equations for the amplitudes
of the quasi-particle excitation
\protect{$\left(u_k(j),v_k(j+1/2)\right)$}.

Solutions are travelling waves partly reflected
and partly transmitted at the interface. In particular,
the incoming and reflected wave \protect{$\left(\Psi_{1L},\Psi_{2L}\right)$}
is a compatible solution for $j<0$ plus
\protect{$u_k\!\left(0\right)=\Psi_{1L}\left(0\right)$} if
\protect{$\Psi_{1L}(j)=\exp{({\rm i}qaj)}-R_k\exp{(-{\rm i}qaj)}$},
\protect{$\Psi_{2L}(j+1/2)=0$},
\protect{$\omega\!\left(q\right)=eV-2t^{\prime}\cos{\left(qa\right)}-\mu$},
namely the tight-binding solution of the left side metal.
Similarly, the transmitted wave
\begin{equation}
{\Psi_{1R}\!\left(j\right) \choose \Psi_{2R}\!\left(j+1/2\right)}
=T_k{u_k \choose v_k{\rm e}^{{\rm i}ka/2}}{\rm e}^{{\rm i}kaj}
\label{eq:transmitted}
\end{equation}
is a solution of
the bulk FE for $j>0$ only if
\protect{$\omega\!\left(k\right)=\xi_k + E_k -\mu$}, with
$\xi_k=\varepsilon_k/2$, $\varepsilon_k=-2t\cos{(ka)}$,
$E_k=[\xi_k^2+|\Delta_k|^2+|V_k|^2]^{1/2}$,
where $\Delta_k=2\Delta\cos{(ka/2)}$ and
$V_k=2{\rm i}V_{df}\sin{(ka/2)}$ have even and odd parity, respectively,
and \protect{$2\left|u_k\right|^2=1+\xi_k/E_k $},
\protect{$\left|u_k\right|^2 + \left|v_k\right|^2 = 1 $}.
Figure \ref{fig1}(b) shows the quasi-particle band structure
on both sides of the junction when $V_{df}\rightarrow 0$.
Note that the FE gap is indirect and that the
bottom of conduction band is much flatter than the top
of valence band.
\begin{figure}
\centerline{\epsfig{file=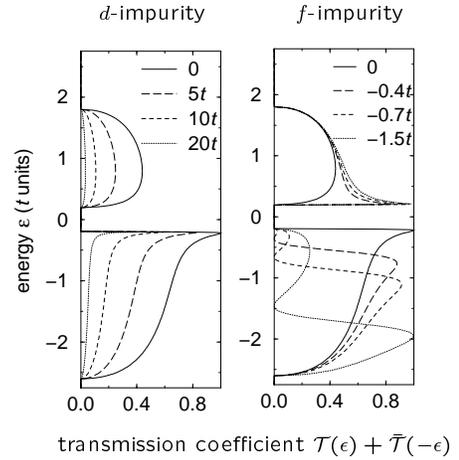,width=2.7in,,angle=90}}
\caption{
\label{fig2}
Total (electron plus hole) transmission coefficient
\protect{${\cal{T}}\!\left(\epsilon\right)+
\bar{\cal{T}}\!\left(-\epsilon\right)$} vs energy, for different
values of impurity levels
$\varepsilon_{d0}$ ($\varepsilon_{f1/2}$).
Left panel: $d$-type impurity. Right panel:
$f$-like impurity.
}
\end{figure}

We now have to match left and right bulk solutions
at the interface.
Because quantities $v_k(1/2)$, $T_k$, and $R_k$ are still unknown,
the finite-difference equations of motion coming
from Eqs.~(\ref{eq:direct}) and (\ref{eq:key})
not yet employed are solved, with the elastic scattering
condition \protect{$\omega(q)=\omega(k)$}.\cite{hirsch}
The reflection coefficient ${\cal{R}}(\omega)$, defined as the ratio
of reflected to incident flux, is simply $\left|R_k\right|^2$
($q,k>0$), and the
transmission coefficient is
\protect{${\cal{T}}\!\left(\omega\right)=
1-{\cal{R}}\!\left(\omega\right)$}.

Electric and heat density currents are calculated to first order in the
differences of $V$ and $T$ on the two sides of the junction.
From these  the conductance \protect{$G=2e^2L_0/h$}, the
thermal conductance
\protect{$G_T=2(L_2-L_1^2/L_0)/(Th)$},
and the thermopower \protect{$Q=L_1/(eTL_0)$} of the interface
are derived,\cite{ziman} where
\begin{equation}
L_n=
\int_{-\infty}^{\infty}\!\!\!{\rm d\,}
\epsilon\left\{{\cal{T}}\!\left(\epsilon\right)+
\bar{{\cal{T}}}\!\left(-\epsilon\right)\right\}\epsilon^n
\left[-\frac{\partial\,f\!\left(\epsilon\right)}
{\partial\,\epsilon}\right],
\label{eq:ldef}
\end{equation}
$f\!\left(x\right)$ is the Fermi distribution function,
and ${\cal{T}}\!\left(x\right)$ [$\bar{{\cal{T}}}\!\left(x\right)$]
is the electron (hole) transmission coefficient,
non-zero only for positive excitation energies within the band range.
The interface figure of merit is $ZT=Q^2GT/G_T$.

Figure \ref{fig2} shows the total transmission coefficient
\protect{${\cal{T}}\!\left(\epsilon\right)+
\bar{{\cal{T}}}\!\left(-\epsilon\right)$} vs energy
for different values of impurity levels $\varepsilon_{d0}$ (left panel) and
$\varepsilon_{f1/2}$ (right panel). We consider a system at $T=0$ with the
same parameters of Fig.~\ref{fig1}(b), i.e.~one in which the metal band is
much
broader than the semiconductor one
and the FE band gap is 0.4$t$,
approximately one tenth of the total bandwidth
(\protect{$V_{df}\rightarrow 0$}). One sees that the clean interface
($\varepsilon_{d0}=\varepsilon_{f1/2}=0$) already presents a strong
electron-hole asymmetry. Hence, from Eq.~(\ref{eq:ldef})
high values of $Q$ follow. The effect of an impurity $d$-layer is just
to uniformly depress ${\cal{T}}+\bar{\cal{T}}$: the greater
the impurity level energy, the lower
the transmission (left panel). This trend, whose
global effect is found to enhance $ZT$, has intrinsic physical limitations
because $\varepsilon_{d0}$ must be of the same order of magnitude as
bulk energies in realistic systems. The behavior of an $f$-impurity is
drastically different. We see that (right panel),
as we set $\varepsilon_{f1/2}$ to negative values, $\bar{\cal{T}}$ goes
to zero in the neighborhood of the same values, as if the hole were
completely backscattered from the interface when resonating with the impurity
atom. A similar
behavior occurs for $\varepsilon_{f1/2}>0$.
The overall effect is so strong even to change the dominant
(electron or hole) character of transport and, hence, the sign of $Q$.
\begin{figure}
\centerline{\epsfig{file=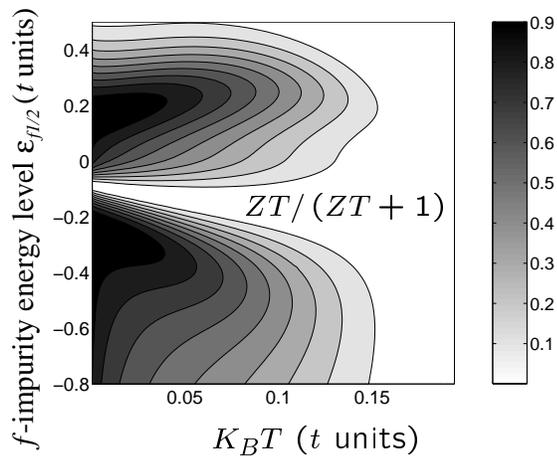,width=2.7in,,angle=90}}
\caption{
\label{fig3}
Two-dimensional contour plot of $ZT/(ZT+1)$ in the $T$-$\varepsilon_{f1/2}$
plane. The gray scale goes from 0
(white) to 1 (black),
and the contour lines are incremented by 0.1.
Parameters as in Fig.~\ref{fig1}.
}
\end{figure}

The effect of the $f$-impurity layer on $ZT$ is shown in Fig.~\ref{fig3}.
We represent the quantity $ZT/(ZT+1)$ (due to
the huge range of $ZT$) as a contour plot
in the $T$-$\varepsilon_{f1/2}$ plane. While $ZT$ has no upper bound,
$ZT/(ZT+1)\leq 1$. For a fixed value of $\varepsilon_{f1/2}$,
$ZT$ goes to zero as $T$ approaches the critical temperature $T_c$
($K_BT_c=0.19t$ here)
at which the gap vanishes and the bulk FE
turns into a metal [\protect{${\cal{T}}\!\left(\epsilon\right)
=\bar{{\cal{T}}}\!\left(-\epsilon\right)$}].
In the opposite limit $T\rightarrow 0$, $ZT$ can reach very
high values
(the darkest region enclosed by the inner contour line corresponds to
$ZT> 90$), depending on $\varepsilon_{f1/2}$.
At $T>T_c$ $ZT$ goes to zero, as $\varepsilon_{f1/2}$
varies, when $Q$ changes sign.

From Fig.~\ref{fig3},
there is an optimal value of $\varepsilon_{f1/2}$ of the interafce layer at
each temperature.  The corresponding maxmimum figure of merit $ZT$
increases with lower temperature.
For example, with a gap of 100 meV at $T=0$, the
best value of $ZT$ will be $\sim\,$1 at $T=$300 K, but
already $\sim\,$6 at 150 K and $\sim\,$100 at 40 K.

In summary, we have made a qualitative theoretical study of the possibility
that a junction of metal and FE (as opposed to
bulk materials) produces high thermopower. The figure of merit $ZT$
attained is very high, especially at low $T$.
In these regimes, bulk thermal conductivity would be dominated by phonons
which would reduce $ZT$. However, if we choose two
materials with large thermal impedance mismatch, the phonon scatterings
at the interface would decrease the junction thermal
conductivity.\cite{lattice}
Thus, phonon conductivity would not
vitiate the high $ZT$ found.
In a longer paper, we will consider the role of phonons in greater detail
and include a more comprehensive study of junctions with
different classes of semiconducting materials.

Work supported by the NSF contract DMR 9721444, INFM PRA99-SSQI,
and Progetto Giovani Ricercatori. We thank
D.~Chemla, M.~L.~Cohen and S.~G.~Louie for the hospitality
at University of California, Berkeley,
where this work was carried out. L.~J.~S.~thanks J.~E.~Hirsch
for helpful discussions.

\clearpage

\end{multicols}

\begin{references}
\bibitem{clarification}
In our calculation $\Delta (T)$ and $\mu (T)$ are bulk FE
values determined self-consistently.
\end{references}
\end{document}